\pdfoutput=1
\documentclass[abstract,headings=normal,DIV=12]{scrartcl}

\usepackage[automark]{scrpage2}
\usepackage[english]{babel}
\usepackage[T1]{fontenc}
\usepackage[applemac]{inputenc}
\usepackage{graphicx}
\usepackage{tikz}
\usepackage{stmaryrd}
\usepackage{amsmath}
\usepackage{amssymb}
\usepackage{amsthm}
\usepackage{subfig}
\usepackage{accents}
\usepackage{url}
\usepackage{booktabs}
\usepackage[colorlinks=true,linkcolor=blue,citecolor=red,urlcolor=black]{hyperref}

\usetikzlibrary{arrows}

\addtokomafont{caption}{\small}

\newcommand{\R}{\mathbb{R}}
\newcommand{\Z}{\mathbb{Z}}

\newcommand{\Ell}{\mathcal{L}}
\newcommand{\ELL}{\mathfrak{L}}

\newcommand{\Es}{\mathcal{S}}
\newcommand{\E}{\mathcal{E}}

\newcommand{\X}{\mathcal{X}}

\renewcommand{\tilde}[1]{\widetilde{#1}}

\makeatletter
\renewcommand*\env@cases[1][1.2]{%
  \let\@ifnextchar\new@ifnextchar
  \left\lbrace
  \def\arraystretch{#1}%
  \array{@{}l@{\quad}l@{}}%
}
\makeatother

\newtheorem{theo}{Theorem}[section]
\newtheorem{lemma}[theo]{Lemma}

\theoremstyle{definition}
\newtheorem{defi}[theo]{Definition}
\newtheorem{ex}[theo]{Example}
\theoremstyle{remark}

\graphicspath{{./graphics/}}
\DeclareGraphicsExtensions{.pdf}

\begin{document}

\title{Two-dimensional variational systems\\ on the root lattice $Q(A_{N})$}
\author{Raphael~Boll}
\publishers{\vspace{0.5cm}{\small Institut f\"ur Mathematik, MA 7-2, Technische Universit\"at Berlin,\\
Stra{\ss}e des 17. Juni 136, 10623 Berlin, Germany\\
E-mail: \href{mailto:boll@math.tu-berlin.de}{\normalfont \ttfamily boll@math.tu-berlin.de}}}
\maketitle

\begin{abstract}
\noindent We study certain two-dimensional variational systems, namely pluri-Lagrangian systems on the root lattice $Q(A_{N})$. Here, we follow the scheme which was already used to define two-dimensional pluri-Lagrangian systems on the lattice $\Z^{N}$ and three-dimensional pluri-Lagrangian systems on the lattice $\Z^{N}$ as well as on $Q(A_{N})$. We will show that the two-dimensional pluri-Lagragian systems on $Q(A_{N})$ are more general than the ones on $\Z^{N}$, in the sense that they can encode several different pluri-Lagrangian systems on $\Z^{N}$. This also means that the variational formulation of several systems of certain hyperbolic equations, so-called quad-equations, can be obtained from one and the same pluri-Lagrangian system on $Q(A_{N})$.\par
\vspace{0,5cm}
\end{abstract}

\section{Introduction}\label{sec:intro}
A variational (or Lagrangian) formulation for certain two-dimensional discrete hyperbolic equations (so-called quad-equations) was established in several papers: While in \cite{ABS1} the action on the two-dimensional lattice $\Z^{2}$ and the corresponding Euler-Lagrange equations were given, in the pioneering work \cite{LN}, the introduction of the concept of discrete 2-forms allowed to consider the action and the corresponding Euler-Lagrange equations on arbitrary 2-manifolds in a higher-dimensional lattice $\Z^{N}$. This concept was applied to the complete list of quad-equations in the well-known ABS classification \cite{ABS1} in \cite{BS1} and for the asymmetric extension~\cite{classification} in \cite{lagrangian}.\par
The weakness of this variational formulation is that the corresponding Euler-Lagrange equations do not coincide with the hyperbolic equations: the former are just consequences of the latter ones or, in other words, the set of solutions of the Euler-Lagrange equations is essentially bigger than the one of the corresponding hyperbolic equations.\par
Similarly, a variational formulation for the three-dimensional discrete KP equation was given in~\cite{LNQ}. In this case, we were able to prove (see \cite{DKP}) that the Euler-Lagrange equations are equivalent to the corresponding hyperbolic equations. Moreover, we also presented a variational formulation with similar properties of the discrete KP equation on the root lattice $Q(A_{N})$ in the same paper.\par
Recently, also the Euler-Lagrange equations considered as integrable systems themselves came more and more into the focus of interest. The underlying theory of pluri-Lagrangian problems was developed for two-dimensional systems in \cite{variational,octahedron}. While in the first of these publications the concept of so-called corner equations as elementary building blocks of Euler-Lagrange equations was introduced, in the second one, the notion of consistency for these corner corner equations was further elucidated. The theory of three-dimensional pluri-Lagrangian problems was developed in \cite{DKP}.\par
In the present paper we will introduce the theory of two-dimensional pluri-Lagrangian systems on the root lattice $Q(A_{N})$ and we will show that these are more general than the ones on the lattices $\Z^{N}$ in the sense that they can encode several different pluri-Lagrangian systems on $\Z^{N}$. This means also that the variational formulation for several systems can be obtained in a straightforward procedure from one and the same pluri-Lagrangian system on $Q(A_{N})$. Moreover, the latter fact corresponds to the fact that certain systems of quad-equations are connected to each other by a certain flipping procedure which was described in~\cite{classification}.\par
Let us start with some concrete definitions valid for an arbitrary $N$-dimensional lattice $\X$.
\begin{defi}[Discrete 2-form]
A \emph{discrete 2-form} on $\X$ is a real-valued function $\Ell$ of oriented 2-cells~$\sigma$ depending on some field $x:\X\to \R$, such that $\Ell$ changes the sign by changing the orientation of~$\sigma$.
\end{defi}
For instance, in $Q(A_{N})$, the 2-cells are triangles, and, in $\Z^{N}$, the 2-cells are quadrilaterals.
\begin{defi}[2-dimensional pluri-Lagrangian problem]
Let $\Ell$ be a discrete 2-form on $\X$ depending on $x:\X\to\R$.
\begin{itemize}
\item To an arbitrary 2-manifold $\Sigma\subset\X$, i.e., a union of oriented 2-cells which forms an oriented two-dimensional topological manifold, there corresponds the \emph{action functional}, which assigns to $x|_{V(\Sigma)}$, i.e., to the fields in the set of the vertices $V(\Sigma)$ of $\Sigma$, the number
\begin{equation*}
\Es_{\Sigma}:=\sum_{\sigma\in\Sigma}\Ell(\sigma).
\end{equation*}
\item We say that the field $x:V(\Sigma)\to \R$ is a critical point of $\Es_{\Sigma}$, if at any interior point $n\in V(\Sigma)$, we have
\begin{equation}\label{eq: dEL gen}
\frac{\partial \Es_{\Sigma}}{\partial x(n)}=0.
\end{equation}
Equations \eqref{eq: dEL gen} are called \emph{discrete Euler-Lagrange equations} for the action $\Es_{\Sigma}$.
\item We say that the field $x:\X\to\R$ solves the \emph{pluri-Lagrangian problem} for the Lagrangian 2-form $\Ell$ if, \emph{for any 2-manifold $\Sigma\subset\X$}, the restriction $x|_{V(\Sigma)}$ is a critical point of the corresponding action $\Es_{\Sigma}$.
\end{itemize}
\end{defi}
\subsection*{Main results}
The main results of the present paper may be summarized as follows:
\begin{itemize}
\item Similar to the case of the lattice $\Z^{N}$, we see how Euler-Lagrange equations on $Q(A_{N})$ arising from pluri-Lagrangian problems form integrable systems on every 2-manifold in the lattice. These are called \emph{pluri-Lagrangian systems}.
\item Pluri-Lagrangian systems on $Q(A_{N})$ can encode several pluri-Lagrangian systems on $\Z^{N}$.
\item The variational formulation of systems of quad-equations which are related by a certain flipping procedure described in~\cite{classification} can be obtained from one and the same pluri-Lagrangian system on $Q(A_{N})$.
\end{itemize}\par
The paper is organized as follows: In Section~\ref{sect:root}, we will give a brief introduction to the theory of the root lattice $Q(A_{N})$ and prove that any 2-manifold with exactly one interior point and be build up as a sum certain elementary building blocks, so-called 3D corners. Then we will study the properties of the discrete 2-forms and the corresponding Euler-Lagrange equations on the root lattice $Q(A_{N})$ in Section~\ref{sect:2form}. After that we will study the relation between the lattices $Q(A_{N})$ and $\Z^{N}$ in Section~\ref{sect:cube} and we will transfer our results from Section~\ref{sect:2form} to the lattice $\Z^{N}$ in Section~\ref{sect:2form2}. In Section~\ref{sect:quad}, we will review the variational formulations of quad-equations. Finally, we will consider a less symmetric example in Section~\ref{sect:asym} which demonstrates that pluri-Lagrangian systems are more general than the ones on $\Z^{N}$ and can encode the variational formulation of several systems of quad-equations.

\section{The root lattice \texorpdfstring{$Q(A_{N})$}{Q(A\_N)}}\label{sect:root}
We consider the root lattice
\[
Q(A_{N}):=\{n:=(n_{0},n_{1},\ldots,n_{N})\in\Z^{N+1}:n_{0}+n_{1}+\ldots+n_{N}=0\},
\]
where $N\geq2$. The two-dimensional sub-lattices $Q(A_{2})$ are given by
\[
Q(A_{2}):=\{(n_{i},n_{j},n_{k}):n_{i}+n_{j}+n_{k}=\mathrm{const}\}.
\]\par
We consider fields $x:Q(A_{N})\to\R$, and use the shorthand notations
\[
x_{\bar{\imath}}=x(n-e_{i}),\qquad x=x(n),\qquad\text{and}\qquad x_{i}=x(n+e_{i}),
\]
where $e_{i}$ is the unit vector in the $i$\textsuperscript{th} coordinate direction.
Furthermore, the shift functions $T_{i}$ and $T_{\bar{\imath}}$ are defined by
\[
T_{i}x_{\alpha}:=x_{i\alpha}\quad\text{and}\quad T_{\bar{\imath}}x_{\alpha}:=x_{\bar{\imath}\alpha}
\]
for a multiindex $\alpha$. For simplicity, we sometimes abuse notations by identifying lattice points $n$ with the corresponding fields $x(n)$.\par
We now give a very brief introduction to the Delaunay cell structure of the $N$-dimensional root lattice $Q(A_{N})$~\cite{CS,MP}. Here, we restrict ourselves to a very elementary description which is appropriate to our purposes and follow the considerations in \cite{ABS3,DKP}. For each $d\leq N$ there are $d$ sorts of $d$-cells of $Q(A_{N})$ denoted by $P(k,d)$ with $k=1,\ldots,d$:
\begin{itemize}
\item One sort of 1-cells:\ \\
\begin{tabular}{ll}\addlinespace
$P(1,1)$:& edges $[ij]:=\{x_{i},x_{j}\};$
\end{tabular}
\item Two sorts of 2-cells:\ \\
\begin{tabular}{ll}\addlinespace
$P(1,2)$:& black triangles $\lfloor ijk\rfloor:=\{x_{i},x_{j},x_{k}\}$;\\\addlinespace
$P(2,2)$:& white triangles $\lceil ijk\rceil:=\{x_{ij},x_{ik},x_{jk}\}$;
\end{tabular}
\item Three sorts of 3-cells:\ \\
\begin{tabular}{ll}\addlinespace
$P(1,3)$:& black tetrahedra $\lfloor ijk\ell\rfloor:=\{x_{i},x_{j},x_{k},x_{\ell }\}$;\\\addlinespace
$P(2,3)$:& octahedra $[ijk\ell]:=\{x_{ij},x_{ik},x_{i\ell},x_{jk},x_{j\ell},x_{k\ell}\}$;\\\addlinespace
$P(3,3)$:& white tetrahedra $\lceil ijk\ell\rceil:=\{x_{ijk},x_{ij\ell},x_{ik\ell},x_{jk\ell}\}$;
\end{tabular}
\bigskip
\end{itemize}
The facets of 2-cells and 3-cells can be found in Appendix~\ref{sec:facets}.\par
In the present paper we will consider objects on \emph{oriented} manifolds. We say that an elementary cell is positively oriented if the indices in the bracket are increasingly ordered. For instance, the black triangle $\lfloor ijk\rfloor$ and white triangle $\lceil ijk\rceil$ are positively oriented if $i<j<k$ (see Figure~\ref{fig:orientation}). Any permutation of two indices changes the orientation to the opposite one.\par
\begin{figure}[htbp]
   \centering
   \subfloat[]{\label{fig:black triangle}
   \begin{tikzpicture}[scale=0.85,inner sep=2]
      \node (x) at (0,0) [circle,fill,label=-135:$x_{i}$] {};
      \node (x1) at (3,0) [circle,fill,label=-45:$x_{j}$] {};
      \node (x3) at (0,3) [circle,fill,label=135:$x_{k}$] {};
      \draw (x) to (x1);
      \draw (x) to (x3);
      \draw (x1) to (x3);
      \draw [ultra thick,->] (0.38,0.88) arc (180:510:0.5);
   \end{tikzpicture}
   }\qquad
   \subfloat[]{\label{fig:white triangle}
      \begin{tikzpicture}[scale=0.85,inner sep=2]
      \node (x12) at (4,1) [circle,fill,label=-45:$x_{ij}$] {};
      \node (x23) at (1,4) [circle,fill,label=135:$x_{jk}$] {};
      \node (x123) at (4,4) [circle,fill,label=45:$x_{ik}$] {};
      \draw (x12) to (x123);
      \draw (x12) to (x23);
      \draw (x23) to (x123);
      \draw [ultra thick,->] (2.62,3.12) arc (180:510:0.5);
   \end{tikzpicture}
   }
   \caption{Orientation of triangles: \protect\subref{fig:black triangle} the black triangle $\lfloor ijk\rfloor$; \protect\subref{fig:white triangle} the white triangle $-\lceil ijk\rceil$}
   \label{fig:orientation}
\end{figure}
When we use the bracket notation, we always write the letters in brackets in increasing order, so, e.g., in writing $\lfloor ijk\rfloor$ we assume that $i<j<k$ and avoid the notation $\lfloor jik\rfloor$ or $\lfloor ikj\rfloor$ for the negatively oriented triangle $-\lfloor ijk\rfloor$.\par
There is a simple recipe to derive the orientation of facets of an $d$-cell: On every index in the brackets we put alternately a ``$+$'' or a ``$-$'' starting with a ``$+$'' on the last index. Then the orientation of each of its facets is indicated by the sign corresponding to the index in the bracket which is omitted. For instance, the octahedron
{
\setlength\arraycolsep{0pt}
\renewcommand{\arraystretch}{0.5}
\begin{equation*}
\begin{array}{cccccc}
&-&+&-&+&\\
{[}&i&j&k&\ell&{]}
\end{array}
\end{equation*}
has the four black triangular facets} $T_{\ell}\lfloor ijk\rfloor$, $-T_{k}\lfloor ij\ell\rfloor$, $T_{j}\lfloor ik\ell\rfloor$, and $-T_{i}\lfloor jk\ell\rfloor$.\par
The following two definitions are valid for arbitrary $N$-dimensional lattices $\X$.
\begin{defi}[Adjacent $d$-cell]
Given an $d$-cell $\sigma$, another $d$-cell $\bar{\sigma}$ is called \emph{adjacent} to $\sigma$ if $\sigma$ and $\bar{\sigma}$ share a common $(d-1)$-cell. The orientation of this $(d-1)$-cell in $\sigma$ must be opposite to its orientation in $\bar{\sigma}$.
\end{defi}
The latter property guarantees that the orientations of the adjacent $d$-cells agree. For instance, in any 2-manifold, any two 2-cells sharing an edge have to be adjacent. Examples for open 2-manifolds in $Q(A_{N})$ are the two-dimensional sub-lattices $Q(A_{2})$, whereas examples of closed 2-manifolds in $Q(A_{N})$ are the set of facets of a tetrahedron (consisting of four triangles of the same color) and the set of facets of an octahedron (consisting of four black and four white triangles).\par
\begin{defi}[Flower]\label{def:flower}
A 2-manifold in $\X$ with exactly one interior vertex $x$ is called a \emph{flower} with center $x$. \emph{The flower} at an interior vertex $x$ of a given 2-manifold is the flower with center $x$ which lies completely in the 2-manifold.
\end{defi}
As a consequence, in $Q(A_{N})$, in each flower every triangle has exactly two adjacent triangles. For instance, in an open 2-manifold $Q(A_{N})$ the flower at an interior vertex consists of six triangles (three black and three white ones).\par
The elementary building blocks of flowers are so-called 3D~corners:
\begin{defi}[3D~corner]
A \emph{3D~corner} with center $x$ is a 2-manifold consisting of all facets of a 3-cell adjacent to $x$.
\end{defi}
In $Q(A_{N})$, there are two different types of 3D corners: a corner on a tetrahedron (consisting of a three triangles of the same color) and a corner on an octahedron (consisting of two black and two white triangles), see Appendix~\ref{sec:corners} for details.\par
The following combinatorial statement will be proven in Appendix~\ref{sec:proof}:
\begin{theo}\label{th:flower}
The flower at any interior vertex of any 2-manifold in $Q(A_{N})$ can be represented as a sum of 3D~corners in $Q(A_{N+2})$.
\end{theo}

\section{Discrete 2-forms on \texorpdfstring{$Q(A_{N})$}{Q(A\_N)}}\label{sect:2form}
Let $\Ell$ be a discrete 2-form on $Q(A_{N})$. The \emph{exterior derivative} $d\Ell$ is a discrete 3-form whose value at any 3-cell in $Q(A_{N})$ is the action functional of $\Ell$ on the 2-manifold consisting of the facets of the 3-cell. For our purposes, we consider discrete 2-forms $\Ell$ vanishing on all white triangles. In particular, we have
\[
d\Ell(\lceil ijk\ell\rceil)\equiv 0
\]
since a white tetrahedron $\lceil ijk\ell\rceil$ only contain white triangles.\par
Moreover, motivated from the case of the lattice $\Z^{N}$ (see~\cite{LN,BS1,lagrangian,variational}), we restrict ourselves to 2-forms of the form
\begin{equation}\label{eq:2-form}
\Ell(\lfloor ijk\rfloor)=\Lambda^{ij}([ij])+\Lambda^{ik}(-[ik])+\Lambda^{jk}([jk])
\end{equation}
on black triangles $\lfloor ijk\rfloor$, where the functions $\Lambda^{ij}$ assigned to the edges $[ij]$ of the triangle are discrete 1-forms which can be different for different lattice directions $i,j$. Therefore, the exterior derivative on a black tetrahedron $\lfloor ijk\ell\rfloor$ is given by
\begin{equation*}
\begin{aligned}
d\Ell(\lfloor ijk\ell\rfloor)&=\Ell(\lfloor ijk\rfloor )+\Ell(-\lfloor ij\ell\rfloor )+\Ell(\lfloor ik\ell\rfloor )+\Ell(-\lfloor jk\ell\rfloor )\\
&=\Lambda^{ij}([ij])+\Lambda^{ik}(-[ik])+\Lambda^{jk}([jk])\\
&\qquad+\Lambda^{ij}(-[ij])+\Lambda^{i\ell}([i\ell])+\Lambda^{j\ell}(-[j\ell])\\
&\qquad+\Lambda^{ik}([ik])+\Lambda^{i\ell}(-[i\ell])+\Lambda^{k\ell}([k\ell])\\
&\qquad+\Lambda^{jk}(-[jk])+\Lambda^{j\ell}([j\ell])+\Lambda^{k\ell}(-[k\ell])\\
&\equiv0
\end{aligned}
\end{equation*}
and the exterior derivative on an octahedron $[ijk\ell]$ is given by
\begin{equation}\label{eq:extdev}
\begin{aligned}
\Es^{ijk\ell}&:=d\Ell([ijk\ell])\\
&\phantom{:}=\Ell(T_{\ell}\lfloor ijk\rfloor)+\Ell(-T_{k}\lfloor ij\ell\rfloor)+\Ell(T_{j}\lfloor ik\ell\rfloor)+\Ell(-T_{i}\lfloor jk\ell\rfloor)\\
&\phantom{:}=\Lambda^{ij}(T_{\ell}[ij])+\Lambda^{ik}(-T_{\ell}[ik])+\Lambda^{jk}(T_{\ell}[jk])\\
&\qquad+\Lambda^{ij}(-T_{k}[ij])+\Lambda^{i\ell}(T_{k}[i\ell])+\Lambda^{j\ell}(-T_{k}[j\ell])\\
&\qquad+\Lambda^{ik}(T_{j}[ik])+\Lambda^{i\ell}(-T_{j}[i\ell])+\Lambda^{k\ell}(T_{j}[k\ell])\\
&\qquad+\Lambda^{jk}(-T_{i}[jk])+\Lambda^{j\ell}(T_{i}[j\ell])+\Lambda^{k\ell}(-T_{i}[k\ell]).
\end{aligned}
\end{equation}
Accordingly, the Euler-Lagrange equations on an octahedron $[ijk\ell]$ are
\begin{equation}\label{eq:EL}
\begin{alignedat}{3}
\frac{\partial \Es^{ijk\ell}}{\partial x_{ij}}&=0,\quad&
\frac{\partial \Es^{ijk\ell}}{\partial x_{ik}}&=0,\quad&
\frac{\partial \Es^{ijk\ell}}{\partial x_{i\ell}}&=0,\\
\frac{\partial \Es^{ijk\ell}}{\partial x_{jk}}&=0,\quad&
\frac{\partial \Es^{ijk\ell}}{\partial x_{j\ell}}&=0,\quad&
\frac{\partial \Es^{ijk\ell}}{\partial x_{k\ell}}&=0.
\end{alignedat}
\end{equation}
These equations are called \emph{corner equations}. The system of six corner equations on an octahedron $[ijk\ell]$ is called \emph{consistent} if it has the minimal possible rank 2, i.e., exactly two equations are independent.\par
The following statement is an immediate consequence of Theorem~\ref{th:flower}:
\begin{theo}\label{cor:Corner}
For every discrete 2-form on $Q(A_{N})$ and every 2-manifold in $Q(A_{N})$ all corresponding Euler-Lagrange equations can be written as sums of corner equations.
\end{theo}

It is obvious that $\Es^{ijk\ell}$ is constant on solutions of \eqref{eq:EL}. The vanishing of this constant, i.e.~the \emph{closedness} of the 2-form, can be seen as a criterion of integrability for pluri-Lagrangian systems (see~\cite{variational,rel} for more explanations).\par
For discrete 2-forms of the form described in \eqref{eq:2-form} the corner equations have a four-leg structure. For instance, the first one in~\eqref{eq:EL} reads as
\begin{equation}\label{eq:EL1}
E_{ij}:=\frac{\partial \Es^{ijk\ell}}{\partial x_{ij}}=\frac{\partial\Lambda^{ik}(T_{j}[ik])}{\partial x_{ij}}+\frac{\partial\Lambda^{i\ell}(-T_{j}[i\ell])}{\partial x_{ij}}+\frac{\partial\Lambda^{jk}(-T_{i}[jk])}{\partial x_{ij}}+\frac{\partial\Lambda^{j\ell}(T_{i}[j\ell])}{\partial x_{ij}}=0.
\end{equation}
We call this equation the \emph{corner equation centered in $x_{ij}$}.\par
\begin{ex}\label{ex:cr}
Consider the discrete 2-form $\Ell$, where for all lattice directions $i,j$ the discrete 1-form $\Lambda^{ij}$ is given by
\[
\Lambda^{ij}([ij]):=(\alpha^{i}-\alpha^{j})\log|x_{i}-x_{j}|.
\]
Here, $\alpha^{i},\alpha^{j}\in\R$ are real parameters assigned to the lattice directions. Then the system~\eqref{eq:EL} of corner equations is consistent and equation~\eqref{eq:EL1} reads as
\[
\frac{\alpha^{i}-\alpha^{k}}{x_{ij}-x_{jk}}-\frac{\alpha^{i}-\alpha^{\ell}}{x_{ij}-x_{j\ell}}-\frac{\alpha^{j}-\alpha^{k}}{x_{ij}-x_{ik}}+\frac{\alpha^{j}-\alpha^{\ell}}{x_{ij}-x_{i\ell}}=0.
\]
For the proof of the consistency of~\eqref{eq:EL} as well as the closedness of the corresponding 2-form is closed on its solutions, we refer to Example~\ref{ex:cr1}.
\end{ex}

\section{The lattice \texorpdfstring{$\Z^{N}$}{Z\^{}N}}\label{sect:cube}
We will now consider the relation between the elementary cells of the root lattice $Q(A_{N})$ and the lattice $\Z^{N}$. The points of $Q(A_{N})$ and of $\Z^{N}$ are in a one-to-one correspondence via
\[
P_{i}:Q(A_{N})\to\Z^{N},\quad x(n_{0},\ldots,n_{i-1},n_{i},n_{i+1},\ldots,n_{N})\mapsto x(n_{0},\ldots,n_{i-1},n_{i+1},\ldots,n_{N}).
\]\par
The two-dimensional elementary cells of $\Z^{N}$ are oriented quadrilaterals
\[
\{jk\}:=\{x,x_{j},x_{k},x_{jk}\}.
\]
We say that the quadrilaral $\{jk\}$ is positively oriented if $j<k$. The permutation of the two indices would change the orientation to the opposite one. Also in this case, we always write the letters in the brackets in increasing order, so, e.g., in writing $\{jk\}$ we assume that $j<k$ and avoid the notation $\{kj\}$ for the negatively oriented quadrilateral $-\{jk\}$.\par
The object in $Q(A_{N})$ which corresponds to the quadrilateral $\{jk\}$ is the sum of two adjacent 2-cells, namely
\begin{itemize}
\item the black triangle $T_{i}\lfloor ijk\rfloor$ (see Figure~\ref{fig:2-cells}\subref{fig:black triangle1}),
\item and the white triangle $-\lceil ijk\rceil$ (see Figure~\ref{fig:2-cells}\subref{fig:white triangle1}).
\end{itemize}
Here, the map $P_{i}$ reads as follows:
\[
x_{ii}\mapsto x,\quad x_{ij}\mapsto x_{j},\quad\text{and}\quad x_{jk}\mapsto x_{jk}.
\]\par
\begin{figure}[htb]
   \centering
   \subfloat[]{\label{fig:black triangle1}
   \begin{tikzpicture}[scale=0.85,inner sep=2]
      \node (x) at (0,0) [circle,fill,label=-135:$x_{ii}$] {};
      \node (x1) at (3,0) [circle,fill,label=-45:$x_{ij}$] {};
      \node (x3) at (0,3) [circle,fill,label=135:$x_{ik}$] {};
      \draw (x) to (x1);
      \draw (x) to (x3);
      \draw (x1) to (x3);
      \draw [ultra thick,->] (0.38,0.88) arc (180:510:0.5);
   \end{tikzpicture}
   }\qquad
   \subfloat[]{\label{fig:white triangle1}
      \begin{tikzpicture}[scale=0.85,inner sep=2]
      \node (x12) at (4,1) [circle,fill,label=-45:$x_{ij}$] {};
      \node (x23) at (1,4) [circle,fill,label=135:$x_{ik}$] {};
      \node (x123) at (4,4) [circle,fill,label=45:$x_{jk}$] {};
      \draw (x12) to (x123);
      \draw (x12) to (x23);
      \draw (x23) to (x123);
      \draw [ultra thick,->] (2.62,3.12) arc (180:510:0.5);
   \end{tikzpicture}
   }\\
   \subfloat[]{\label{fig:quad}
   \begin{tikzpicture}[scale=0.85,inner sep=2]
      \node (x) at (0,0) [circle,fill,label=-135:$x_{ii}$] {};
      \node (x1) at (3,0) [circle,fill,label=-45:$x_{ij}$] {};
      \node (x3) at (0,3) [circle,fill,label=135:$x_{ik}$] {};
      \node (x13) at (3,3) [circle,fill,label=45:$x_{jk}$] {};
      \draw (x) to (x1);
      \draw (x) to (x3);
      \draw (x1) to (x13);
      \draw (x3) to (x13);
      \draw [ultra thick,->] (1,1.5) arc (180:510:0.5);
   \end{tikzpicture}
   }\qquad
   \subfloat[]{\label{fig:quad1}
   \begin{tikzpicture}[scale=0.85,inner sep=2]
      \node (x) at (0,0) [circle,fill,label=-135:$x$] {};
      \node (x1) at (3,0) [circle,fill,label=-45:$x_{j}$] {};
      \node (x3) at (0,3) [circle,fill,label=135:$x_{k}$] {};
      \node (x13) at (3,3) [circle,fill,label=45:$x_{jk}$] {};
      \draw (x) to (x1);
      \draw (x) to (x3);
      \draw (x1) to (x13);
      \draw (x3) to (x13);
      \draw [ultra thick,->] (1,1.5) arc (180:510:0.5);
   \end{tikzpicture}
   }
   \caption{Two adjacent 2-cells of the lattice $Q(A_{N})$: \protect\subref{fig:black triangle1} black triangle $T_{i}\lfloor ijk\rfloor$, \protect\subref{fig:white triangle1} white triangle $-\lceil ijk\rceil$. The sum \protect\subref{fig:quad} of these 2-cells corresponds to the quadrilateral~$\{jk\}$ (see~\protect\subref{fig:quad1}) in $\Z^{N}$.}
   \label{fig:2-cells}
\end{figure}
As three-dimensional elementary cells of $\Z^{N}$, we consider oriented cubes denoted by
\[
\{jk\ell\}:=\{x,x_{j},x_{k},x_{\ell},x_{jk},x_{j\ell},x_{k\ell},x_{jk\ell}\}.
\]
We say that the cube $\{jk\ell\}$ is positively oriented if $j<k<\ell$. Any permutation of two indices changes the orientation to the opposite one. Also in this case, we always write the letters in the brackets in increasing order, so, e.g., in writing $\{jk\ell\}$ we assume that $j<k<\ell$ and avoid the notation $\{kj\ell\}$ or $\{j\ell k\}$ for the negatively oriented cube $-\{jk\ell\}$.\par
The object in $Q(A_{N})$ which corresponds to the cube $\{jk\ell\}$ is the sum of three adjacent 3-cells, namely
\begin{itemize}
\item the black tetrahedron $-T_{i}\lfloor ijk\ell\rfloor$ (see Figure~\ref{fig:3-cells}\subref{fig:black tetrahedron}),
\item the octahedron $[ijk\ell]$ (see Figure~\ref{fig:3-cells}\subref{fig:octahedron}),
\item and the white tetrahedron $-T_{\bar{\imath}}\lceil ijk\ell\rceil$ (see Figure~\ref{fig:3-cells}\subref{fig:white tetrahedron}).
\end{itemize}
It contains sixteen triangles and to every quadrilateral face of $\{jk\ell\}$ there corresponds a pair of these triangles containing one black and one white triangle. Here, the map $P_{i}$ reads as follows:
\[
x_{ii}\mapsto x,\quad x_{ij}\mapsto x_{j},\quad x_{jk}\mapsto x_{jk},\quad\text{and}\quad x_{\bar{\imath}jk\ell}\mapsto x_{jk\ell}.
\]\par
\begin{figure}[htb]
   \centering
   \subfloat[]{\label{fig:black tetrahedron}
   \begin{tikzpicture}[scale=0.85,inner sep=2]
      \node (x) at (0,0) [circle,fill,label=-135:$x_{ii}$] {};
      \node (x1) at (3,0) [circle,fill,label=-45:$x_{ij}$] {};
      \node (x2) at (1,1) [circle,fill,label=180:$x_{ik}$] {};
      \node (x3) at (0,3) [circle,fill,label=135:$x_{i\ell}$] {};
      \draw (x) to (x1);
      \draw [dashed] (x) to (x2);
      \draw (x) to (x3);
      \draw [dashed] (x1) to (x2);
      \draw (x1) to (x3);
      \draw [dashed] (x2) to (x3);
   \end{tikzpicture}
   }
   \subfloat[]{\label{fig:octahedron}
   \begin{tikzpicture}[scale=0.85,inner sep=2]
      \node (x1) at (3,0) [circle,fill,label=-45:$x_{ij}$] {};
      \node (x2) at (1,1) [circle,fill,label=-135:$x_{ik}$] {};
      \node (x3) at (0,3) [circle,fill,label=135:$x_{i\ell}$] {};
      \node (x12) at (4,1) [circle,fill,label=-45:$x_{jk}$] {};
      \node (x13) at (3,3) [circle,fill,label=45:$x_{j\ell}$] {};
      \node (x23) at (1,4) [circle,fill,label=135:$x_{k\ell}$] {};
      \draw (x1) to (x2);
      \draw (x1) to (x3);
      \draw (x1) to (x12);
      \draw (x1) to (x13);
      \draw (x2) to (x3);
      \draw [dashed] (x2) to (x12);
      \draw [dashed] (x2) to (x23);
      \draw (x3) to (x13);
      \draw (x3) to (x23);
      \draw (x12) to (x13);
      \draw [dashed] (x12) to (x23);
      \draw (x13) to (x23);
   \end{tikzpicture}
   }
   \subfloat[]{\label{fig:white tetrahedron}
      \begin{tikzpicture}[scale=0.85,inner sep=2]
      \node (x12) at (4,1) [circle,fill,label=-45:$x_{jk}$] {};
      \node (x13) at (3,3) [circle,fill,label=0:$x_{j\ell}$] {};
      \node (x23) at (1,4) [circle,fill,label=135:$x_{k\ell}$] {};
      \node (x123) at (4,4) [circle,fill,label=45:$x_{\bar{\imath}jk\ell}$] {};
      \draw (x12) to (x13);
      \draw (x12) to (x123);
      \draw (x12) to (x23);
      \draw (x13) to (x23);
      \draw (x13) to (x123);
      \draw (x23) to (x123);
   \end{tikzpicture}
   }\\
   \subfloat[]{\label{fig:cube}
   \begin{tikzpicture}[scale=0.85,inner sep=2]
      \node (x) at (0,0) [circle,fill,label=-135:$x_{ii}$] {};
      \node (x1) at (3,0) [circle,fill,label=-45:$x_{ij}$] {};
      \node (x2) at (1,1) [circle,fill,label=180:$x_{ik}$] {};
      \node (x3) at (0,3) [circle,fill,label=135:$x_{i\ell}$] {};
      \node (x12) at (4,1) [circle,fill,label=-45:$x_{jk}$] {};
      \node (x13) at (3,3) [circle,fill,label=0:$x_{j\ell}$] {};
      \node (x23) at (1,4) [circle,fill,label=135:$x_{k\ell}$] {};
      \node (x123) at (4,4) [circle,fill,label=45:$x_{\bar{\imath}jk\ell}$] {};
      \draw (x) to (x1);
      \draw [dashed] (x) to (x2);
      \draw (x) to (x3);
      \draw [dashed] (x1) to (x2);
      \draw (x1) to (x3);
      \draw (x1) to (x12);
      \draw (x1) to (x13);
      \draw [dashed] (x2) to (x3);
      \draw [dashed] (x2) to (x12);
      \draw [dashed] (x2) to (x23);
      \draw (x3) to (x13);
      \draw (x3) to (x23);
      \draw (x12) to (x13);
      \draw (x12) to (x123);
      \draw [dashed] (x12) to (x23);
      \draw (x13) to (x23);
      \draw (x13) to (x123);
      \draw (x23) to (x123);
   \end{tikzpicture}
   }\qquad
   \subfloat[]{\label{fig:cube1}
   \begin{tikzpicture}[scale=0.85,inner sep=2]
      \node (x) at (0,0) [circle,fill,label=-135:$x$] {};
      \node (x1) at (3,0) [circle,fill,label=-45:$x_{j}$] {};
      \node (x2) at (1,1) [circle,fill,label=180:$x_{k}$] {};
      \node (x3) at (0,3) [circle,fill,label=135:$x_{\ell}$] {};
      \node (x12) at (4,1) [circle,fill,label=-45:$x_{jk}$] {};
      \node (x13) at (3,3) [circle,fill,label=0:$x_{j\ell}$] {};
      \node (x23) at (1,4) [circle,fill,label=135:$x_{k\ell}$] {};
      \node (x123) at (4,4) [circle,fill,label=45:$x_{jk\ell}$] {};
      \draw (x) to (x1);
      \draw [dashed] (x) to (x2);
      \draw (x) to (x3);
      \draw (x1) to (x12);
      \draw (x1) to (x13);
      \draw [dashed] (x2) to (x12);
      \draw [dashed] (x2) to (x23);
      \draw (x3) to (x13);
      \draw (x3) to (x23);
      \draw (x12) to (x123);
      \draw (x13) to (x123);
      \draw (x23) to (x123);
   \end{tikzpicture}
   }
   \caption{Three adjacent 3-cells of the lattice $Q(A_{N})$: \protect\subref{fig:black tetrahedron} black tetrahedron $-T_{i}\lfloor ijk\ell\rfloor$, \protect\subref{fig:octahedron} octahedron $[ijk\ell]$, \protect\subref{fig:white tetrahedron} white tetrahedron $-T_{\bar{\imath}}\lceil ijk\ell\rceil$. The sum \protect\subref{fig:cube} of these 3-cells corresponds to a cube~\protect\subref{fig:cube1}.}
   \label{fig:3-cells}
\end{figure}
Also in this case there is an easy recipe to obtain the orientation of the facets of an (oriented) cube: on every index between the brackets we put alternately a ``$+$'' and a ``$-$'' starting with a ``$+$'' on the last index. Then we get each facet by deleting one index and putting the corresponding sign in front of the bracket. For instance, the cube
{
\setlength\arraycolsep{0pt}
\renewcommand{\arraystretch}{0.5}
\begin{equation*}
\begin{array}{ccccc}
&+&-&+&\\
\{&j&k&\ell&\}
\end{array}
\end{equation*}
has the six facets: the quadrilaterals $\{jk\}$, $-\{j\ell\}$, $\{k\ell\}$, and the opposite ones $-T_{\ell}\{jk\}$, $T_{k}\{j\ell\}$, and $-T_{j}\{k\ell\}$.\par}
As a consequence of Definition~\ref{def:flower}, in each flower in $\Z^{N}$, every quadrilateral has exactly two adjacent quadrilaterals.\par
For the proof of the following theorem, which is the analogue of Theorem~\ref{cor:Corner}, we refer to~\cite{variational}.
\begin{theo}\label{th:flower cube}
The flower at any interior vertex of any 2-manifold in $\Z^{N}$ can be represented as a sum of 3D~corners in $\Z^{N+1}$.
\end{theo}

\section{Discrete 2-forms on \texorpdfstring{$\Z^{N}$}{Z\^{}N}}\label{sect:2form2}
Let $\ELL$ be a discrete 2-form on $\Z^{N}$. The \emph{exterior derivative} $d\ELL$ is a discrete 3-form whose value at any cube in $\Z^{N}$ is the action functional of $\ELL$ on the 2-manifold consisting of the facets of the cube:
\begin{align*}
\mathfrak{S}^{jk\ell}&:=d\ELL(\{jk\ell\})\\
&\phantom{:}=\ELL(\{jk\})+\ELL(-\{j\ell\})+\ELL(\{k\ell\})+\ELL(-T_{\ell}\{jk\})+\ELL(T_{k}\{j\ell\})+\ELL(-T_{j}\{k\ell\}).
\end{align*}
Accordingly, the Euler-Lagrange equations on the cube $\{jk\ell\}$ are given by
\begin{equation}\label{eq:EL cube}
\begin{alignedat}{6}
\frac{\partial \mathfrak{S}^{jk\ell}}{\partial x}&=0,\\
\frac{\partial \mathfrak{S}^{jk\ell}}{\partial x_{j}}&=0,\quad&
\frac{\partial \mathfrak{S}^{jk\ell}}{\partial x_{k}}&=0,\quad&
\frac{\partial \mathfrak{S}^{jk\ell}}{\partial x_{\ell}}&=0,\\
\frac{\partial \mathfrak{S}^{jk\ell}}{\partial x_{jk}}&=0,\quad&
\frac{\partial \mathfrak{S}^{jk\ell}}{\partial x_{j\ell}}&=0,\quad&
\frac{\partial \mathfrak{S}^{jk\ell}}{\partial x_{k\ell}}&=0,\\
\frac{\partial \mathfrak{S}^{jk\ell}}{\partial x_{jk\ell}}&=0.
\end{alignedat}
\end{equation}
They are called \emph{corner equations}.\par
The following statement is an immediate consequence of Theorem~\ref{th:flower cube}:
\begin{theo}\label{cor:Corner cubic}
For every discrete 2-form on $\Z^{N}$ and every 2-manifold in $\Z^{N}$ all corresponding Euler-Lagrange equations can be written as a sum of corner equations.
\end{theo}
We are interested in discrete 2-forms $\ELL$ defined as
\[
\ELL:=(P_{i})_{\star}\Ell,
\]
where $\Ell$ is a discrete 2-form on the root lattice $Q(A_{N})$ as specified in~\eqref{eq:2-form}. Therefore, $\ELL$ evaluated at the quadrilateral $\{jk\}$ reads as
\begin{align*}
\ELL(\{jk\})&=((P_{i})_{\star}\Ell)(P_{i}(T_{i}\lfloor ijk\rfloor-\lceil ijk\rceil))\\
&=(P_{i})_{\star}(\Ell(T_{i}\lfloor ijk\rfloor)-\underbrace{\Ell(\lceil ijk\rceil)}_{=0})\\
&=(P_{i})_{\star}(\Lambda^{ij}([ij])+\Lambda^{ik}(-[ik])+\Lambda^{jk}([jk])).
\end{align*}\par
For this discrete 2-form, one can see after a simple computation that the exterior derivative is given by
\begin{align*}
\mathfrak{S}^{jk\ell}&=d\ELL(\{jk\ell\})\\
&\phantom{:}=(P_{i})_{\star}(\Lambda^{ij}(T_{\ell}[ij])+\Lambda^{ik}(-T_{\ell}[ik])+\Lambda^{jk}(T_{\ell}[jk])\\
&\qquad\qquad+\Lambda^{ij}(-T_{k}[ij])+\Lambda^{i\ell}(T_{k}[i\ell])+\Lambda^{j\ell}(-T_{k}[j\ell])\\
&\qquad\qquad+\Lambda^{ik}(T_{j}[ik])+\Lambda^{i\ell}(-T_{j}[i\ell])+\Lambda^{k\ell}(T_{j}[k\ell])\\
&\qquad\qquad+\Lambda^{jk}(-T_{i}[jk])+\Lambda^{j\ell}(T_{i}[j\ell])+\Lambda^{k\ell}(-T_{i}[k\ell]))\\
&\phantom{:}=(P_{i})_{\star}d\Ell([ijk\ell])
\end{align*}
Therefore, there are no corner equations on the cube $\{jk\ell\}$ centered at $x$ and $x_{jk\ell}$ since $\mathfrak{S}^{jk\ell}$ does not depend on these two variables. The remaining corner equations from~\eqref{eq:EL cube} are given by
\begin{align}\label{eq:EL cube1}
\E_{j}&:=\frac{\partial \mathfrak{S}^{jk\ell}}{\partial x_{j}}=(P_{i})_{\star}\frac{\partial \Es^{ijk\ell}}{\partial x_{ij}}=(P_{i})_{\star}E_{ij}\\
\intertext{and}\label{eq:EL cube2}
\E_{k\ell}&:=\frac{\partial \mathfrak{S}^{jk\ell}}{\partial x_{k\ell}}=(P_{i})_{\star}\frac{\partial \Es^{ijk\ell}}{\partial x_{k\ell}}=(P_{i})_{\star}E_{k\ell}.
\end{align}\par
The system of six corner equations on a cube $\{jk\ell\}$ is called \emph{consistent} if it has the minimal possible rank 2, i.e., exactly two equations are independent.\par
Also in this case, it is obvious that $\mathfrak{S}^{jk\ell}$ is constant on solutions of \eqref{eq:EL cube1}, \eqref{eq:EL cube2}. Again, the vanishing of this constant, i.e.~the \emph{closedness} of the 2-form, can be seen as a criterion of integrability for pluri-Lagrangian systems (see~\cite{variational,rel} for more explanations).\par

\begin{ex}\label{ex:cr1}
Consider the 2-form $\ELL=(P_{i})_{\star}\Ell$, where the 1-forms $\Lambda^{ij}$ are defined as in Example~\ref{ex:cr}, i.e.,
\[
\Lambda^{ij}([ij]):=(\alpha^{i}-\alpha^{j})\log|x_{i}-x_{j}|
\]
with $\alpha^{i},\alpha^{j}\in\R$. Then equations~\eqref{eq:EL cube1} and \eqref{eq:EL cube2} read as
\begin{align}\label{eq:EL cube3}
&\frac{\alpha^{i}-\alpha^{k}}{x_{j}-x_{jk}}-\frac{\alpha^{i}-\alpha^{\ell}}{x_{j}-x_{j\ell}}-\frac{\alpha^{j}-\alpha^{k}}{x_{j}-x_{k}}+\frac{\alpha^{j}-\alpha^{\ell}}{x_{j}-x_{\ell}}=0\\
\intertext{and}\label{eq:EL cube4}
&\frac{\alpha^{i}-\alpha^{k}}{x_{\ell}-x_{k\ell}}-\frac{\alpha^{j}-\alpha^{k}}{x_{j\ell}-x_{k\ell}}-\frac{\alpha^{i}-\alpha^{\ell}}{x_{k}-x_{k\ell}}+\frac{\alpha^{j}-\alpha^{\ell}}{x_{jk}-x_{k\ell}}=0.
\end{align}
For the proof of the consistency of the system of corner equations as well as of the closedness of the corresponding 2-form, we refer to~\cite{variational,octahedron}. Obviously, the consistency of this system of corner equations and the closedness of the corresponding 2-form are equivalent to those in Example~\ref{ex:cr}.
\end{ex}

\section{Quad-equations}\label{sect:quad}
On the lattice $\Z^{N}$ we consider certain hyperbolic systems, so-called \emph{quad-equations}. These are equations of the form
\[
Q_{jk}(x,x_{j},x_{k},x_{jk})=0,
\]
where $Q_{jk}\in\R[x,x_{j},x_{k},x_{jk}]$ is an irreducible multi-affine polynomial. The system of the six (possibly different) quad-equations
\begin{alignat}{2}\label{eq:quad}
&Q_{jk}(x,x_{j},x_{k},x_{jk})=0,&\qquad
&\bar{Q}_{jk}(x_{\ell},x_{j\ell},x_{k\ell},x_{jk\ell})=0
\end{alignat}
assigned to the faces of the cube $\{jk\ell\}$ is called \emph{consistent} if it has the minimal possible rank~$4$, i.e., exactly four equations are independent. For classification of consistent systems of quad-equations we refer to~\cite{ABS1,classification}.\par
A variational formulation for quad-equations was given in~\cite{ABS1,LN,BS1,lagrangian} in the following sense: For a system of quad-equation there is given a discrete 2-form such that every solution of the system of quad-equations satisfies the system of the corresponding corner equations but not vice versa. This can be seen in a very transparent way: one can write every corner equation as a difference of two quad-equations on adjacent quadrilaterals in the so-called three-leg form.
\begin{ex}\label{ex:cr2}
Consider the discrete 2-form defined in Example~\ref{ex:cr} and the corresponding corner equations~\eqref{eq:EL cube3} and \eqref{eq:EL cube4}. Since only differences of $\alpha$s appear, one can set $\alpha^{i}=0$ without restriction. Then one can write the equations~\eqref{eq:EL cube3} and \eqref{eq:EL cube4} as
\begin{align*}
&\left(\frac{\alpha^{j}}{x_{j}-x}-\frac{\alpha^{k}}{x_{j}-x_{jk}}-\frac{\alpha^{j}-\alpha^{k}}{x_{j}-x_{k}}\right)-\left(\frac{\alpha^{j}}{x_{j}-x}-\frac{\alpha^{\ell}}{x_{j}-x_{j\ell}}-\frac{\alpha^{j}-\alpha^{\ell}}{x_{j}-x_{\ell}}\right)=0\\
\intertext{and}
&\left(\frac{\alpha^{j}}{x_{jk\ell}-x_{k\ell}}-\frac{\alpha^{k}}{x_{\ell}-x_{k\ell}}-\frac{\alpha^{j}-\alpha^{k}}{x_{j\ell}-x_{k\ell}}\right)-\left(\frac{\alpha^{j}}{x_{jk\ell}-x_{k\ell}}-\frac{\alpha^{\ell}}{x_{k}-x_{k\ell}}-\frac{\alpha^{j}-\alpha^{\ell}}{x_{jk}-x_{k\ell}}\right)=0,
\end{align*}
i.e., they can be written as differences of two three-leg forms.\par
The system of three-leg forms arising by this procedure applied to all corner equations in $\Z^{N}$ is equivalent to the system of quad-equations on $\Z^{N}$, where each quadrilateral $\{jk\}$ carry a copy
\begin{equation*}
\frac{x-x_{j}}{x-x_{k}}\cdot\frac{x_{k}-x_{jk}}{x_{j}-x_{jk}}=\frac{\alpha^{j}}{\alpha^{k}}
\end{equation*}
of the so-called cross-ratio equation ($Q_{1}^{0}$ following the notation of the classification in \cite{classification}).\par
Therefore, every solution of the system of cross-ratio equations solves the system of corner equations, but there are solutions of the system of corner equations which do not solve the system of cross-ratio equations.
\end{ex}

\section{A less symmetric example}\label{sect:asym}
In the example of the variational formulation for the cross-ratio equation, which we considered in Examples~\ref{ex:cr}, \ref{ex:cr1}, and \ref{ex:cr2}, all 1-forms $\Lambda^{ij}$ coincide up to the parameters $\alpha^{i}$ and $\alpha^{j}$. Therefore, different maps $P_{i}$ lead to the same discrete 2-form on the lattice $\Z^{N}$ up to a different enumeration of lattice directions. However, as already mentioned, one can choose 1-forms which are different for different lattice directions. Then different maps $P_{i}$ lead to essentially different discrete 2-form on $\Z^{N}$. In other words, there are different 2-forms on $\Z^{N}$ which can be encoded in one and the same discrete 2-form on the root lattice $Q(A_{N})$. We will demonstrate this with a simple example.\par
We consider the discrete 2-form
\[
\Ell(\lfloor ijk\rfloor)=\Lambda^{ij}([ij])+\Lambda^{ik}(-[ik])+\Lambda^{jk}([jk])
\]
on $Q(A_{3})$, where
\begin{alignat*}{2}
\Lambda^{01}([01])&:=-x_{0}x_{1},&\qquad
\Lambda^{23}([23])&:=(\alpha^{2}-\alpha^{3})\log|x_{2}-x_{3}|,\\
\Lambda^{02}([02])&:=-x_{0}x_{2},&\qquad
\Lambda^{13}([13])&:=(\alpha^{1}-\alpha^{3})\log|x_{1}-x_{3}|,\\
\Lambda^{03}([03])&:=-x_{0}x_{3},&\qquad
\Lambda^{12}([12])&:=(\alpha^{1}-\alpha^{2})\log|x_{1}-x_{2}|.
\end{alignat*}
Then the corner equations on the octahedron $[0123]$ are given by
\begin{align*}
E_{0j}&:=-x_{jk}+x_{j\ell}-\frac{\alpha^{j}-\alpha^{k}}{x_{0j}-x_{0k}}+\frac{\alpha^{j}-\alpha^{\ell}}{x_{0j}-x_{0\ell}}=0,\\
E_{k\ell}&:=x_{0\ell}-\frac{\alpha^{j}-\alpha^{k}}{x_{j\ell}-x_{k\ell}}-x_{0k}+\frac{\alpha^{j}-\alpha^{\ell}}{x_{jk}-x_{k\ell}}=0,
\end{align*}
where $j,k,\ell\in\{1,2,3\}$ and $j\neq k\neq\ell\neq j$.\par
On the one hand, one can consider the 2-form $\ELL:=(P_{0})_{\star}\Ell$ on $\Z^{N}$ as in the example before. Then the corner equations on the cube $\{123\}$ are given by
\begin{align*}
\E_{j}&:=-x_{jk}+x_{j\ell}-\frac{\alpha^{j}-\alpha^{k}}{x_{j}-x_{k}}+\frac{\alpha^{j}-\alpha^{\ell}}{x_{j}-x_{\ell}}=0,\\
\E_{k\ell}&:=x_{\ell}-\frac{\alpha^{j}-\alpha^{k}}{x_{j\ell}-x_{k\ell}}-x_{k}+\frac{\alpha^{j}-\alpha^{\ell}}{x_{jk}-x_{k\ell}}=0,
\end{align*}
where $j,k,\ell\in\{1,2,3\}$ and $j\neq k\neq\ell\neq j$.\par
Again, the corner equations can be written as differences of three-leg forms:
\begin{align*}
\E_{j}&=\left(x-x_{jk}-\frac{\alpha^{j}-\alpha^{k}}{x_{j}-x_{k}}\right)-\left(x-x_{j\ell}-\frac{\alpha^{j}-\alpha^{\ell}}{x_{j}-x_{\ell}}\right)=0,\\
\E_{k\ell}&=\left(x_{\ell}-x_{jk\ell}-\frac{\alpha^{j}-\alpha^{k}}{x_{j\ell}-x_{k\ell}}\right)-\left(x_{k}-x_{jk\ell}-\frac{\alpha^{j}-\alpha^{\ell}}{x_{jk}-x_{k\ell}}\right)=0.
\end{align*}
The system of three-leg forms is equivalent to the system of quad-equations, where all equations are copies of the discrete KdV equation (rhombic-type equation $H_{1}^{0}$ in the notation of the classification in \cite{classification}), namely
\begin{equation}\label{eq:KdV}
(x-x_{jk})(x_{j}-x_{k})=\alpha^{j}-\alpha^{k},
\end{equation}
where $j,k\in\{1,2,3\}$ and $j\neq k$, assigned to the quadrilaterals of the lattice such that, in every cube, opposite quadrilaterals carry the same equation.\par
On the other hand, one can consider the 2-form $\tilde{\ELL}:=(P_{3})_{\star}\Ell$ on $\Z^{N}$. Then the corner equations on the cube $\{012\}$ are given by
\begin{align*}
\tilde\E_{0}&:=-x_{1}+x_{2}-\frac{\alpha^{1}-\alpha^{3}}{x_{01}-x_{0}}+\frac{\alpha^{2}-\alpha^{3}}{x_{02}-x_{0}}=0,\\
\tilde\E_{1}&:=-x_{0}+\frac{\alpha^{1}-\alpha^{2}}{x_{1}-x_{2}}+x_{01}-\frac{\alpha^{2}-\alpha^{3}}{x_{12}-x_{1}}=0,\\
\tilde\E_{2}&:=x_{0}-\frac{\alpha^{1}-\alpha^{2}}{x_{1}-x_{2}}-x_{02}+\frac{\alpha^{1}-\alpha^{3}}{x_{12}-x_{2}}=0,\\
\tilde\E_{01}&:=-x_{12}+x_{1}-\frac{\alpha^{1}-\alpha^{2}}{x_{01}-x_{02}}+\frac{\alpha^{1}-\alpha^{3}}{x_{01}-x_{0}}=0,\\
\tilde\E_{02}&:=x_{12}-x_{2}+\frac{\alpha^{1}-\alpha^{2}}{x_{01}-x_{02}}-\frac{\alpha^{2}-\alpha^{3}}{x_{02}-x_{0}}=0,\\
\tilde\E_{12}&:=x_{02}-\frac{\alpha^{1}-\alpha^{3}}{x_{12}-x_{2}}-x_{01}+\frac{\alpha^{2}-\alpha^{3}}{x_{12}-x_{1}}=0.
\end{align*}
Since only differences of $\alpha$s are involved, one can set $\alpha^{3}=0$ without restriction. Then one can write the corner equations as differences of three-leg forms:
\begin{align*}
\tilde\E_{0}&:=\left(x-x_{1}-\frac{\alpha^{1}}{x_{01}-x_{0}}\right)-\left(x-x_{2}-\frac{\alpha^{2}}{x_{02}-x_{0}}\right)=0,\\
\tilde\E_{1}&:=\left(-x_{0}+x_{01}-\frac{\alpha^{1}}{x-x_{1}}\right)-\left(-\frac{\alpha^{1}}{x-x_{1}}+\frac{\alpha^{2}}{x_{12}-x_{1}}-\frac{\alpha^{1}-\alpha^{2}}{x_{1}-x_{2}}\right)=0,\\
\tilde\E_{2}&:=\left(x_{0}-x_{02}+\frac{\alpha^{2}}{x-x_{2}}\right)-\left(\frac{\alpha^{2}}{x-x_{2}}-\frac{\alpha^{1}}{x_{12}-x_{2}}+\frac{\alpha^{1}-\alpha^{2}}{x_{1}-x_{2}}\right)=0,\\
\tilde\E_{01}&:=\left(-x_{12}+x_{1}+\frac{\alpha^{2}}{x_{01}-x_{012}}\right)-\left(\frac{\alpha^{2}}{x_{01}-x_{012}}-\frac{\alpha^{1}}{x_{01}-x_{0}}+\frac{\alpha^{1}-\alpha^{2}}{x_{01}-x_{02}}\right)=0,\\
\tilde\E_{02}&:=\left(x_{12}-x_{2}-\frac{\alpha^{1}}{x_{02}-x_{012}}\right)-\left(-\frac{\alpha^{1}}{x_{02}-x_{012}}+\frac{\alpha^{2}}{x_{02}-x_{0}}-\frac{\alpha^{1}-\alpha^{2}}{x_{01}-x_{02}}\right)=0,\\
\tilde\E_{12}&:=\left(x_{02}-x_{012}-\frac{\alpha^{1}}{x_{12}-x_{2}}\right)-\left(x_{01}-x_{012}-\frac{\alpha^{2}}{x_{12}-x_{1}}\right)=0.
\end{align*}
The system of three-leg forms is equivalent to the system
\begin{equation}\label{eq:asym}
\begin{alignedat}{3}
&(x-x_{1})(x_{0}-x_{01})=-\alpha^{1},&\qquad
&(x-x_{2})(x_{0}-x_{02})=-\alpha^{2},&\qquad
&\frac{x-x_{1}}{x-x_{2}}\cdot\frac{x_{2}-x_{12}}{x_{1}-x_{12}}=\frac{\alpha^{1}}{\alpha^{2}}
\end{alignedat}
\end{equation}
of quad-equations. Also in this case, in every cube, opposite quadrilaterals carry the same equations. Four of them are trapezoidal-type equations $H_{1}^{0}$, the remaining two are copies of the cross-ratio equation ($Q_{1}^{0}$ in the notation of \cite{classification}).\par
So, the variational formulation of two essentially different systems of quad-equations is encoded in one and the same discrete 2-form $\Ell$ on the root lattice $Q(A_{N})$.\par
Even if the systems~\eqref{eq:KdV} and \eqref{eq:asym} are different, they are known to be related by a certain flipping procedure which was already mentioned in \cite{classification} which we will repeat here. The systems of quad-equations from the (extended) ABS-list \cite{ABS1,classification} possess the \emph{tetrahedron property}, i.e., for every solution of the system of quad-equations~\eqref{eq:quad} on the faces of a cube two further quad-equations, so called \emph{tetrahedron equations},
\begin{align*}
&T(x,x_{jk},x_{j\ell},x_{k\ell})=0,\\
\intertext{and}
&\bar{T}(x_{j},x_{k},x_{\ell},x_{jk\ell})=0
\end{align*}
are satisfied automatically. Now applying the flipping procedure to \eqref{eq:quad}, i.e. interchanging the variables $x_{k}$ and $x_{jk}$ as well as $x_{\ell}$ and $x_{j\ell}$, we get a new system of quad-equations
\begin{equation*}
\begin{alignedat}{2}
&Q_{jk}(x,x_{j},x_{jk},x_{k})=0,&\qquad
&\bar{Q}_{jk}(x_{j\ell},x_{\ell},x_{k\ell},x_{jk\ell})=0,\\
&Q_{j\ell}(x,x_{j},x_{j\ell},x_{\ell})=0,&\qquad
&\bar{Q}_{j\ell}(x_{jk},x_{k},x_{k\ell},x_{jk\ell})=0,\\
&T(x,x_{jk},x_{j\ell},x_{k\ell})=0,&\qquad
&\bar{T}(x_{j},x_{k},x_{\ell},x_{jk\ell})=0
\end{alignedat}
\end{equation*}
on the cube which is proven to be 3D consistent in \cite{classification}.

\section{Conclusion}
We come to the conclusion that it is more natural to consider two-dimensional pluri-Lagrangian systems on the lattice $Q(A_{N})$ rather than on $\Z^{N}$. Having experienced this also in the three-dimensional case (see \cite{DKP}) gives us reason to believe that this is always to be considered.\par
In~\cite{variational}, we pointed out that corner equations on $\Z^{N}$ are more general than quad-equations: every solution of the system of quad-equations satisfies the system of corner equations, but generic solutions of the system of corner equations do not satisfy the system of quad-equations. In the present paper, we showed that corner equations on $Q(A_{N})$ are even more general: there are essentially different systems of corner equations on $\Z^{N}$ which can be encoded in one and the same system of corner equations on $Q(A_{N})$. Consequently, there are systems of corner equations on $Q(A_{N})$ that encodes the variational formulation for essentially different systems of quad-equations on $\Z^{N}$.\par
In the present paper, we only illustrated the theory by using two examples. However, in an analogous way, we can construct many more examples by adjusting the discrete 2-forms coming from the variational formulation~\cite{LN,BS1,lagrangian} of quad-equations of the (extended) ABS-list~\cite{ABS1,classification}.

\section*{Acknowledgments}
This research was supported by the DFG Collaborative Research Center TRR 109 ``Discretization in Geometry and Dynamics''.

\appendix
\section{Facets of \texorpdfstring{$d$}{d}-cells of the root lattice \texorpdfstring{$Q(A_{N})$}{Q(A\_N)}}\label{sec:facets}
\paragraph{Facets of 2-cells:}\ \\
\begin{tabular}{ll}\addlinespace
Black triangles $\lfloor ijk\rfloor$: & three edges $[ij]$, $-[ik]$, and $[jk]$;\\\addlinespace
White triangles $\lceil ijk\rceil$: & three edges $T_{k}[ij]$, $-T_{j}[ik]$, and $T_{i}[jk]$;
\end{tabular}
\paragraph{Facets of 3-cells:}\ \\
\begin{tabular}{ll}\addlinespace
Black tetrahedra $\lfloor ijk\ell\rfloor$:& four black triangles $\lfloor ijk\rfloor$, $-\lfloor ij\ell\rfloor$, $\lfloor ik\ell\rfloor$, and $-\lfloor jk\ell\rfloor$;\\\addlinespace
Octahedra $[ijk\ell]$:& four black triangles $T_{\ell}\lfloor ijk\rfloor$, $-T_{k}\lfloor ij\ell\rfloor$, $T_{j}\lfloor ik\ell\rfloor$, and $-T_{i}\lfloor jk\ell\rfloor$,\\
& four white triangles $\lceil ijk\rceil$, $-\lceil ij\ell\rceil$, $\lceil ik\ell\rceil$, and $-\lceil jk\ell\rceil$;\\\addlinespace
White tetrahedra $\lceil ijk\ell\rceil$:& four white triangles $T_{\ell}\lceil ijk\rceil$, $-T_{k}\lceil ij\ell\rceil$, $T_{j}\lceil ik\ell\rceil$, and $-T_{i}\lceil jk\ell\rceil$;
\end{tabular}

\section{3D corners on 3-cells of the root lattice \texorpdfstring{$Q(A_{N})$}{Q(A\_N)}}\label{sec:corners}
\paragraph{Black tetrahedron $\lfloor ijk\ell\rfloor$:} The 3D corner with center vertex $x_{i}$ contains
\begin{itemize}
\item the three black triangles $\lfloor ijk\rfloor$, $-\lfloor ij\ell\rfloor$, and $\lfloor ik\ell\rfloor$;
\end{itemize}
\paragraph{Octahedron $[ijk\ell]$:} The 3D~corner with center vertex $x_{ij}$ contains
\begin{itemize}
\item the two black triangles $T_{j}\lfloor ik\ell\rfloor$, and $-T_{i}\lfloor jk\ell\rfloor$,
\item and the two white triangles $\lceil ijk\rceil$, and $-\lceil ij\ell\rceil$;
\end{itemize}
\paragraph{White tetrahdedron $\lceil ijk\ell\rceil$:} The 3D~corner with center vertex $x_{ijk}$ contains
\begin{itemize}
\item the three white triangles $-T_{k}\lceil ij\ell\rceil$, $T_{j}\lceil ik\ell\rceil$, and $-T_{i}\lceil jk\ell\rceil$.
\end{itemize}

\section{Proof of Theorem~\ref{th:flower}}\label{sec:proof}
Set $M:=N+1$ and $L:=N+2$. Then, for the construction of the sum $\Sigma$ of 3D~corners representing the flower $\sigma$ centered in $X$, we use the following algorithm:
\begin{enumerate}
\item For every black triangle $\pm\lfloor ijk\rfloor\in\sigma$ at the interior vertex $X$ we add the 3D~corner with center vertex $X$ on the black tetrahedron $\pm\lfloor ijk M\rfloor$ to $\Sigma$.
\item For every white triangle $\pm\lceil ijk\rceil\in\sigma$ we add the 3D~corner with center vertex $X$ on the octahedron $\pm[ijk M]$ to $\Sigma$.
\item For every white triangle $\pm\lceil ijM\rceil\in\Sigma\setminus\sigma$ which appeared in $\Sigma$ during the previous step we add the 3D~corner with center vertex $X$ on the white tetrahedron $\mp T_{\bar{L}}\llceil ijML\rrceil$ to $\Sigma$.
\end{enumerate}
Therefore, we have to prove that $\Sigma=\sigma$.\par
Assume that $X=x_{i}$. Then for each black triangle $\pm\lfloor ijk\rfloor\in\sigma$ we added the two black triangles $\mp\lfloor ijM\rfloor$ and $\pm\lfloor ik M\rfloor$ to $\Sigma$ which do not belong to $\sigma$. Moreover, $\pm\lfloor ijk\rfloor$ has two facets adjacent to $x_{i}$, namely $\pm[ij]$, which is the common edge with $\mp\lfloor ijM\rfloor$ (up to orientation), and $\pm [ik]$, which is the common edge with $\mp\lfloor ik\ell M\rfloor$. Therefore, each of these black triangles has to cancel away with the corresponding black triangle from the 3D~corner which is coming from the triangle adjacent to $\pm\lfloor ijk\rfloor$ via the corresponding edge.\par
Assume that $X=x_{ij}$. Then for each white triangle $\pm\lceil ijk\rceil\in\sigma$ we added the two black triangles $\pm T_{j}\lfloor ik M\rfloor$, and $\mp T_{i}[jkM]$ as well as the white triangle $\mp\lceil ijM\rceil$ to $\Sigma$ which do not belong to $\sigma$. Moreover, $\pm\lceil ijk\rceil$ has two facets adjacent to $x_{ij}$, namely $\mp T_{j}[ik]$, which is the common edge with $\pm T_{j}[ikM]$, and $\mp T_{i} [jk]$, which is the common edge with $\mp T_{i}[jkM]$. Therefore, each of these black triangles has to cancel away with the corresponding black triangle from the 3D~corner which is coming from the triangle adjacent to $\pm\lceil ijk\rceil$ via the corresponding edge.\par
Consider two triangles $\Omega,\bar{\Omega}\in\sigma$ adjacent via the edge $ [ij]$, say $[ij]$ belongs to $\Omega$ and $-[ij]$ belongs to $\bar{\Omega}$. Then the 3D~corner corresponding to $\Omega$ contributes the black triangle $-\lfloor ijM\rfloor$ to $\Sigma$, whereas the 3D~corner corresponding to $\bar{\Omega}$ contributes the black triangle $\lfloor ijM\rfloor$ to $\Sigma$. Therefore, the latter two black tetrahedra cancel out.\par
Up to know we proved that all black triangles in $\Sigma\setminus\sigma$ cancel out. We will now continue with the white triangles in $\Sigma\setminus\sigma$.\par
\begin{lemma}
The white tetrahedra $\lfloor ijM\rfloor$ arising in the second step of the algorithm build flowers which only contain white triangles.
\begin{proof}
We have two prove that each of these white triangles has exactly two adjacent white triangles in the flowers, one via each edge adjacent to $X$. They are not adjacent to the triangles in $\sigma$, but each of them has two common neighbors adjacent to $X$ with the corresponding white triangle in $\sigma$. These common neighbors are black triangles which cancel out in the previous steps of the algorithm.\par
Consider now two white triangles $T,\bar{T}\in\Sigma\setminus\sigma$, where the corresponding white triangles in $\sigma$ are adjacent, i.e., there is a pair of black triangles with the same set of points and different orientation, one adjacent to $T$ and the other adjacent to $\bar{T}$. Therefore, $T$ and $\bar{T}$ share a common edge (up to orientation), i.e., they are adjacent.\par
Consider the 3D corner which we add to $\Sigma$ for a black triangle. Its two black triangles which do not belong to $\sigma$ share a common edge (up to orientation) whose points do no lie in $\sigma$. Furthermore, consider a sequence of adjacent black triangles in $\sigma$. Then the black triangles in the corresponding 3D~corners which are not in $\sigma$ all share a common edge (up to orientation).\par
Consider now two white triangles $T,\bar{T}\in\Sigma\setminus\sigma$, where the corresponding triangles in $\sigma$ are connected by a sequence of black triangles in $\sigma$. Then, there is a pair of black triangles, one of them adjacent to $T$ and the other one adjacent to $\bar{T}$, which share a common edge (up to orientation). This edge does not belong to any triangle in $\sigma$ and, therefore, is a common edge of $T$ and $\bar{T}$, i.e., $T$ and $\bar{T}$ are adjacent.
\end{proof}
\end{lemma}
Now we continue with the proof of Theorem~\ref{th:flower}. We already proved that a flower containing only black triangles can be written as a sum of 3D corners on black tetrahedra (see proof of step 1). Analogously, one can write every flower containing only white tetrahedra as a sum of 3D corners on white tetrahedra. So we write for each of the flowers of white tetrahedra in $\Sigma\setminus\sigma$ after the second step of the algorithm the flower of opposite orientation as a sum of 3D corners on white tetrahedra and add this sum to $\Sigma$. Then $\Sigma=\sigma$.
\qed

{\small
\bibliographystyle{amsalpha}
\bibliography{Quellen}
}

\end{document}